\documentclass[11pt,twoside]{article}

\usepackage{asp2004}
\usepackage{epsf}
\usepackage{lscape}
\usepackage{graphicx}

\begin{document}
\title{Extending the redshift distribution of submm galaxies:  Identification of a $z\approx4$ submm galaxy}   %%% Fill in title
\author{K.K.~Knudsen$^1$, J.-P.~Kneib$^{2,3}$, E.~Egami$^4$}   %%% Fill in author names
\affil{$^1$ Max-Planck-Institut f\"ur Astronomie, K\"onigstuhl 17, D-69117 Heidelberg, Germany\\ 
$^2$ OAMP, Laboratoire d'Astrophysique de Marseille, Traverse du Siphon, 13012 Marseille, France\\ 
$^3$ California Institute of Technology, MS 105-24, Pasadena, CA 91125\\
$^4$ Steward Observatory, University of Arizona, 933 North Cherry Avenue, Tucson, AZ 85721}    %%% Fill in author affiliations

\begin{abstract} %%% Abstract to run on from here.
We present the identification of a bright submillimeter galaxy (SMG) in 
the field of Abell\,2218.  The galaxy has a spectroscopic redshift of 
$\sim$\,4, and is currently the highest redshift SMG known.  It is 
detected at all 
wavelengths from optical to submm, including the {\it Spitzer} IRAC bands. 
We discuss the properties of this galaxy, which is undergoing intense 
star formation at a rate $\sim$\,600\,M$_\odot$yr$^{-1}$.  We also compare 
the properties to those of radio-preselected submm-bright galaxies.  
The $z\approx4$ result extends the redshift distribution of SMGs.
\end{abstract}

%%% MAIN BODY OF TEXT GOES HERE. CONSULT "INSTRUCTIONS FOR AUTHORS USING
%%% LATEX2E MARKUP", SECTIONS 2.3-2.6 FOR HELP WITH EQUATIONS, FIGURES,
%%% AND TABLES.

%\section{}   %%% Top level section head (remove "%" symbol)
%\subsection{}   %%% Second level section head (remove "%" symbol)
%\subsubsection{}   %%% Lowest level section head (remove "%" symbol)
%\section*{}	%%% Unnumbered top level section head (remove "%" symbol)
%\subsection*{}   %%% Unnumbered second level section head (remove "%" symbol)

\section{Introduction}

Major progress in submillimeter cosmology has been seen since the 
commissioning of SCUBA (Submillimetre Common-User Bolometer Array; 
Holland et al. 1999), which was mounted on the James Clerk Maxwell 
Telescope (JCMT).   Many studies have been undertaken studying 
submillimeter galaxies (SMGs) in large blank field surveys and surveys 
of strongly lensing galaxy clusters.  In the cluster surveys the 
gravitational lensing moves the confusion limit to flux levels fainter 
than the blank field limit of 2 mJy.   It has taken several years to 
overcome the many complications involved in the identification of 
the counterparts.  The large beam of 15$''$ in diameter
 makes unique identifications 
difficult.  However, resorting to multiwavelength follow-up ranging 
from radio to optical it is possible to constrain the properties of 
the underlying galaxy, including (photometric) redshift and the 
presence of AGN. 
(For more details, see Smail 2006, this volume.)

Recently, Chapman et al. (2003, 2005) published a large spectroscopic 
survey of radio-detected bright SMGs ($f_{850} > 4$\,mJy).  They presented 
redshifts for  about 70 SMGs from several blank field surveys --- an 
effort, which has been crucial for quantifying 
the properties of the bright SMGs (see Smail 2006).  The 
redshift distribution has a median redshift of ~2.4.  Even though using 
radio observations to determine the position of the underlying galaxy is 
a very efficient identification technique, it is likely biased towards 
redshifts $z<3.5$.  
There are candidate counterparts towards higher redshifts 
(e.g.\ Dunlop et al., 2004), however, their redshifts still need to be 
confirmed spectroscopically.   We here present details and spectroscopy 
for a redshift 4 counterpart.  

\begin{figure}
\begin{center}
\includegraphics[width=11.0cm]{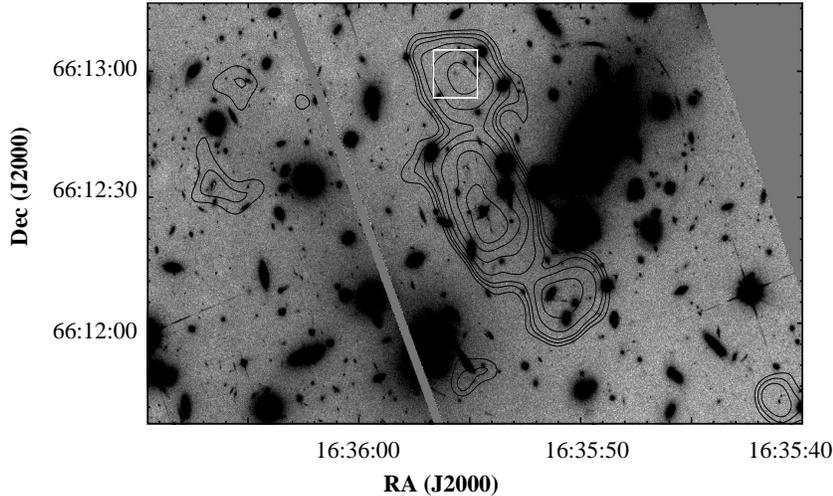}
\caption[]{ACS F850LP image overlayed with the 850\,$\mu$m SCUBA contours 
at $3, 4, 5, 7, 10, 15, 20\sigma$.  
The white box indicate position of SMM\,J16359+66130 and of the sub-image 
shown in Fig.~\ref{fig2}.  The three bright sources south of the box are 
the triple-imaged galaxies SMM\,J16359+6612 (Kneib et al., 2004a). 
\label{fig1}
}
\end{center}
\end{figure}

\section{SMM\,J16359+66130 -- redshift 4}

\begin{figure}
\begin{minipage}{6.5cm}
%\begin{center}
\includegraphics[width=6.5cm]{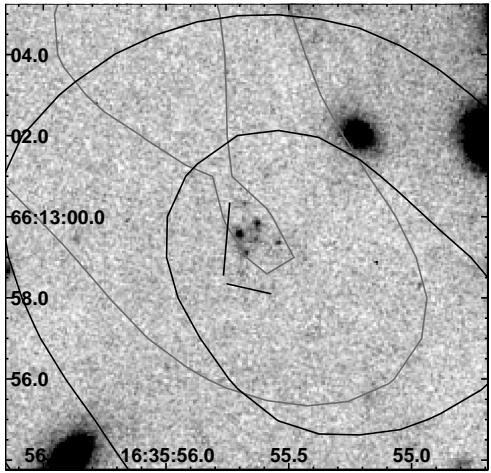}
\end{minipage}
\begin{minipage}{7.4cm}
\vskip.7cm
\caption[]{ACS F850LP sub-image of the $z\approx 4$ source, SMM\,J16359+66130, 
with SCUBA 
contours overlayed (black show 850$\mu$m and grey 450$\mu$m).   
The two solid lines indicate the extent of the galaxy, which corrected 
for lensing corresponds to a physical scale of 6.2kpc$\times$2.8kpc.   
The largest knot has a scale corresponding to $\sim$0.5\,kpc. \label{fig2}}
\end{minipage}
\end{figure}

Deep SCUBA maps at both 850$\mu$m and 450$\mu$m have been obtained of  
the cluster field Abell\,2218 (Knudsen et al., 2006).  Nine sources were 
detected, four of these having 850-fluxes around 10 mJy; 
an optical image with the SCUBA contours is shown in Fig.~\ref{fig1}.  
Three of these bright sources were identified as a multiply imaged background 
galaxy at $z=2.516$ (SMMJ16359+6612; Kneib et al., 2004a).  It is for the 
fourth of the bright sources, SMMJ16359+66130, that we here propose the 
identification (Knudsen, Kneib \& Egami, 2006, in preparation).  

SMMJ16359+66130 is detected both at 850$\mu$m and 450$\mu$m with fluxes 
of 11 mJy and 20 mJy, respectively.  The lensing magnification of the 
source is about 4.5.  The low  flux ratio $f_{450}/f_{850}\sim 2$  is 
strongly suggestive of this SMG being at redshift $z>3.5$.  It is not 
detected in the radio maps at 1.4 GHz and 8.2 GHz from Garrett et al. 
(2005), which supports that this is a high redshift source, $z>2.5$.  

\begin{figure}
\begin{minipage}{8.0cm}
\includegraphics[width=8cm]{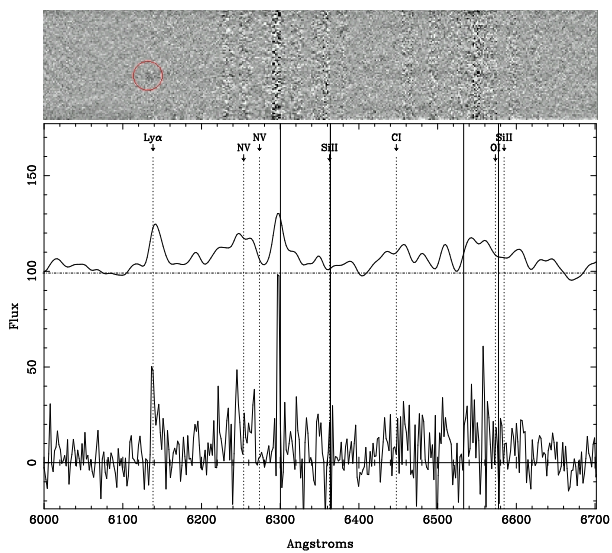}
\end{minipage}
\begin{minipage}{5.9cm}
\vskip-.5cm
\caption[]{
Optical  spectrum of SMMJ16359+66130.   (Top)  The 2D sky subtracted 
spectrum that shows a faint emission line.   (Bottom)  The 1D sky 
subtracted spectrum, for which a smoothed version is displayed at the 
top of the figure.  Solid vertical lines indicate the location of 
bright sky lines. 
We identify the bright asymmetric emission line with Ly$\alpha$ ; this implies 
a redshift $z=4.048$.   
\label{fig3}
}
\end{minipage}
\end{figure}

In the optical and near-infrared images there is only one galaxy within a 
search radius of 8$''$ from the submm position, which is at very high 
redshift.  It is indicated in the optical HST ACS F850LP $z'-$image 
(Kneib et al., 2004b) in Fig.~\ref{fig2}.  
The position is RA,Dec(J2000) = 16:35:55.66,+66:12:59.5 and it has a
magnitude of $z'_{850lp} = 23.90\pm0.04$\,mag and of 
$R_{702w} = 25.4\pm0.12$\,mag, though is undetected in the deep WFPC2 
F606W and F450W images from Smail et al.\ (2001). 
We have obtained optical spectroscopy for this source using LRIS on Keck.   
In the spectrum, which is shown in Fig.~\ref{fig3}, we identify one emission 
line.  The line appears to be asymmetric and with a lower continuum 
level on the blue side.  We propose that this is Ly$\alpha$, which implies a 
redshift of $z=4.048\pm0.003$.

\begin{figure}
\begin{minipage}{8cm}
\includegraphics[width=8cm]{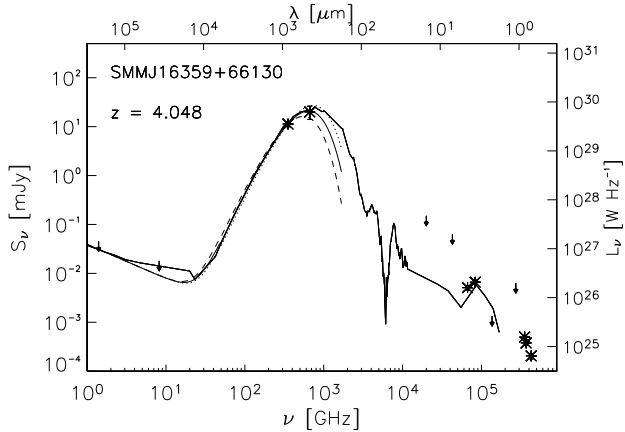}
\end{minipage}
\begin{minipage}{5.9cm}
\caption[]{
Spectral energy distribution ranging from radio to optical with
measurements and $3\sigma$ upper limits  from VLA, WSRT, JCMT (SCUBA), ISO, 
Spitzer (IRAC), WHT/Gemini  and HST.  The fluxes have not been corrected 
for lensing magnification.  The SED is overplotted with the Arp220 
SED and with modified blackbody SEDs (with T=30,35,40K).
\label{fig4}
}
\end{minipage}
\end{figure}

SMMJ16359+66130 is detected at both in the optical and at near-infrared 
wavelengths, including the {\it Spitzer} IRAC images of A2218 
(Egami et al., 2005), where the fluxes are $f_{3.6} = 5.6\pm0.7$\,$\mu$Jy 
and $f_{4.5} = 4.5\pm0.5$\,$\mu$Jy.  
In Fig.~\ref{fig4}, we show the spectral energy distribution. 
The star formation rate is about 600\,M$_\odot$yr$^{-1}$ as estimated 
from the SCUBA detection.   There is no indication of the presence of a 
strong AGN.

\section{Extending the redshift distribution}

Even though the unlensed submm flux is slightly fainter compared to the 
sample of Chapman et al.\ (2003, 2005), the properties of SMMJ16359+66130 
are similar to those of the radio-selected identifications of the 
Chapman et al. sample.  There are no signs of a strong, dominant AGN, 
and so likely the dominant heating mechanism is star formation.  
Additionally, Ly$\alpha$ is seen in emission.  
Hence SMMJ16359+66130 is likely 
similar to many sources in the Chapman et al.\ sample.  

The high redshift of SMMJ16359+66130 extends the redshift distribution 
of SMGs as established by Chapman et al.\ (2003, 2005).  While the highest 
redshift identification in the sample of Chapman et al. is 3.6, they 
present models for the redshift distribution indicating that less than 
10\% of the SMGs would lie beyond that.   
For the current SMG samples of a total of a few hundred sources, one
would expect a few tens of SMGs with $z>4$. 

%\acknowledgements %%% Text of acknowledgements runs on after this command.

%%% THE BIBLIOGRAPHY
%%%
%%% CONSULT SECTION 3 OF "INSTRUCTIONS FOR AUTHORS" FOR HOW TO USE NATBIB.
%%% AUTHORS ARE ENCOURAGED TO USE EITHER THE "THEBIBLIOGRAPY" ENVIRONMENT
%%% BY UNCOMMENTING (DELETING THE "%" SYMBOL) THE COMMANDS BELOW, OR BY
%%% USING THE BIBTEX ENVIRONMENT. TO FIND OUT WHICH IS APPLICABLE TO YOUR
%%% CONTRIBUTION, CONSULT THE VOLUME EDITORS FOR YOUR PROCEEDINGS.
%%%


\begin{thebibliography}{}
\bibitem[Chapman et al.(2003)]{chapman03}Chapman, S.C., et al., 2003, Nature, 422, 695
\bibitem[2005]{chapman05}Chapman, S.C., et al., 2005, ApJ, 622, 772
\bibitem[2004]{dunlop04}Dunlop, J.S., et al., 2004, MNRAS, 350, 769 
\bibitem[2005]{egami05}Egami, E., et al., 2005, ApJ, 618, L5 
\bibitem[2005]{garrett05}Garrett, M., et al., 2005, A\&A, 431, L21
\bibitem[1999]{holland99}Holland, W., et al., 1999, MNRAS, 303, 659
\bibitem[2004a]{kneib04}Kneib J.-P., et al., 2004a, MNRAS, 349, 1211 
\bibitem[2004b]{kneib04b}Kneib J.-P., et al., 2004b, ApJ, 607, 697 
\bibitem[2006]{knudsen06}Knudsen K.K., et al., 2006, MNRAS, in press (astro-ph/0602131)
\bibitem[2001]{smail01}Smail, I., et al., 2001, MNRAS, 323, 839 
\bibitem[2006]{smail06}Smail, I., 2006, this volume 
\end{thebibliography}
\end{document}